# Behavioral subtyping through typed assertions

**Herbert Toth**,   herbert.toth@inode.at

**Abstract**   This paper presents a critical discussion of popular approaches to ensure the Liskov substitution principle in class hierarchies (e.g. Design by Contract™, specification inheritance). It will be shown that they have some deficiencies which are due to the way how effective constraints are calculated for subclass methods. A new mechanism, called *client conformance*, is introduced that takes the client's view on the program state into account more properly: The client's static type determines the context in which reasoning about program state is to be done. This is the context to which the runtime assertion checking (RAC) of server methods must be adapted appropriately. In a stepwise argumentation we show the improvements for RAC that can be reached following this approach in a natural way, preserving the percolation pattern mechanism: Clients will neither be confronted with unsafe or surprising executions, nor with surprising failures of server methods.

**Keywords**   Liskov substitution principle, behavioral subtyping, percolation pattern, Design by Contract™, specification inheritance, client conformance, runtime assertion checking (RAC), unsafe execution, surprising execution.

## 1   Introduction

Beginning with the late eighties and early nineties there has been an increasing interest in the topic of assertion based software development as a possibility to increase the quality of software products. Though the use of pre- and post-conditions to specify software dates back at least to [Hoare69], [Liskov88] and [Meyer92] may be regarded as parents of the many contributions that emerged during the last twenty-five years: Many books and papers (sometimes highly technical as e.g. [Back98] or [LeNa06a, LeNa06b]) have been written about the practical and theoretical challenges of how to use assertions for specifying and checking object-oriented software during daily development. In addition to the theoretical investigations, a considerable number of tools (see e.g. [Plösch02], [Froi07] for Java and [Toth06] for C++) have been presented around the year 2000, implementing the prevailing contemporary knowledge, which - until recently - has almost always been the way how Bertrand Meyer has defined the Design by Contract™ mechanism for Eiffel [Meyer97]. Current research has gone much farther; in fact far beyond what average software developers know, are interested in or get time to study, and which – unfortunately - cannot properly be expected from them. To get an impression of what is going on start out at one of the places given at the footnote of this page[1], to name just a few.

In this paper I will discuss basic concepts and properties of the idea of behavioral subtyping in an informal way. My viewpoint will be that of a practitioner, but with some theoretical underpinnings. In order to get a rather straightforward course of argumentation, I will focus on the fundamental parts of assertion based software development: method specifications and (class, object) invariants. This means a considerable simplification by leaving aside some otherwise important aspects: callbacks and reentrance [Fähn08], multiple inheritance, aliasing problems [Meyer10], and exceptions [SouFri00] will not be considered. But despite these confinements I believe that things remain interesting and complex enough to make the following considerations worth the work. Moreover, it is the *conceptual* foundations that must be valid first and foremost: this is the only way to construct suitable *formal* systems upon a sound basis.

This paper (a substantial revision of [Toth10]) is of a metalevel character: instead of using a running example, possible configurations are described in an abstract way using truth values of constraints only. It is organized as follows: Sections 2 and 3 present the conceptual and formal preliminaries for the subsequent considerations. In section 4 we describe the commonly used approach for the calculation of effective preconditions and the resulting problems, and develop a client conforming alternative. Section 5 does the same for postcondition and invariants. We continue with a short characterization of client conforming specifications in class hierarchies, and prove some desirable properties of client conforming method specifications, and close with putting together all the findings for client conforming RAC. In section 7 we present a detailed analysis of Liskov's Method Rule and give a short

---

[1] `http://research.microsoft.com/en-us/um/people/leino/papers.html`, `http://pm.inf.ethz.ch/publications/index.html`, `http://se.inf.ethz.ch/publications/`, `http://www.eecs.ucf.edu/~leavens/JML/papers.shtml`, `http://www.eecs.ucf.edu/~leavens/main.html`, `http://front.math.ucdavis.edu/cs.LO`, or [HaLeav12]

discussion of some further issues of software contract formulation, and section 8 summarizes the paper. Finally, in the Appendix we try to explain the prevalence of simple percolation and specification inheritance by some important logical relationships that can be shown, and also provide an overview on the relationships between the three approaches.

## 2   Preliminaries, notation and terminology

The technical means for specifying behavioral properties of software are *assertions*: logical predicates over certain domains that, if they should become verifiable by runtime assertion checking (RAC), must be suitably coded as boolean expressions in the implementation (or some annotation) language used for constructing a software system. Assertions are used to set up so-called *contracts* and can be distinguished according to the role they are intended to play in program specifications (we closely follow [Binder00], chapter 17 and [Toth05]).

### 2.1   Software contracts

*Contracts on method level:* It is common practice to say that assertions at the method level define what must hold between the caller of a method and the method called: Assertions of this kind check method calls for proper invocation and method code for correct computation. But be careful: As you will learn soon, what is really checked at runtime on method borders can differ significantly from what is written as method contracts by developers.

- *Preconditions* express the requirements that clients must satisfy whenever they call a method. They are obligations for the client/caller, and benefits for the server/callee.
- *Postconditions* describe what the server guarantees on return, if the method's preconditions have been satisfied on entry: These are the obligations for the server/callee, and benefits for the client/caller.

*Contracts on class level: Invariants* are class-specific predicates that define what states of an object are considered to be consistent by spelling our constraints for the state space of each of a class's instances: They describe all value combinations that instances of that class are allowed to hold in all client-visible states i.e. within the scope of this paper: after creation, and before and after any call of a public method.

*Contracts on class hierarchy level:* At the class hierarchy level, relationships between superclass and subclass specifications can be checked. The LSP approach requires a subclass's contract to be consistent with the contracts of all its superclasses: A subtype must require no more and promise no less than its supertype. This ensures that subclass objects can be substituted for superclass objects without causing failures or requiring special case code on client side. For more discussion see sections 3.2 and 7.1 below.

*Design by Contract™ (DbC)* is a class design technique that has been devised by Bertrand Meyer and has been made an essential part of the Eiffel language (see e.g. [Meyer97]): to make contracts executable is the real breakthrough novelty about DbC. Its paradigm is as follows: A contract is an explicit statement of rights and obligations between a client and a server, and states what both parties must do, independent of how this is accomplished. The client's obligation is to call a method only in a program state where both the class invariant and the method's precondition hold. The method, in turn, guarantees that the work specified in the postcondition has been done, and the class invariant is still respected. A precondition violation thus points out an error on client side, and a postcondition failure a bug in the implementation of the routine. The failure of an invariant can be a bug of either the client or the server, depending on whether it is detected in the context of pre- or postcondition checking, i.e. before or after method execution.

### 2.2   Formal basics

We will often speak about predicates (or Boolean expressions in a programming language), about evaluating them to **true** or **false**, and about operations with and relationships between such expressions. Therefore, for the reader's convenience, we summarize some definitions and theorems from propositional logic used later on in this paper in *Table 1*. Beyond these simple logical laws (simply referred to as (Tnn) in the text), the following terminology and notation will be used in the subsequent considerations:

**Definition 1.** (Strength of predicates). A predicate $A$ is said to be *stronger than* another predicate $B$, if it logically implies it, i.e. $A \rightarrow B$. If $A$ is stronger than $B$, then $B$ is said to be *weaker than* $A$. Thus **false** is the strongest, and **true** is the weakest constraint.

**Notation.** (Classes and subclasses). Let $C$ denote a class; we take $C := S^0C$, and for $k \geq 1$ $S^kC := SS^{k-1}C$. Furthermore we assume that, for $k \geq 1$, $S^kC$ is a subclass of $S^{k-1}C$; thus, *SC* is subclass of *C*, and *SSC* subclass of *SC*. We use *supers(C)* to denote the set of superclasses of *C*, i.e. *supers*($S^kC$) := $\{ S^iC \mid 0 \leq i \leq k \}$. Note, that by this definition a class *C* is a (trivial) superclass of itself.

**Notation.** (Methods and constraints). A method defined in class *C* with name *m* is denoted by $m_C$. For a class *C* we denote its invariant by $inv_C$; the pre- and postcondition for an arbitrary, but otherwise fixed method *m* in class *C* are denoted by $pre_C$ and $post_C$, respectively. Assertions of any kind that are actually checked at runtime will be called *effective constraints*. (Note: Only invariants will be prefixed with **old_** to indicate if its value before method execution is taken, whereas for preconditions this is always implicit.)

| | |
|---|---|
| (T1)  $A \wedge B \rightarrow A$ | (T11) $(A \rightarrow B) \wedge (B \rightarrow C) \rightarrow (A \rightarrow C)$ |
| (T2)  $A \rightarrow A \vee B$ | (T12) $A \wedge (A \rightarrow B) \equiv A \wedge B$ |
| (T3)  **true** $\wedge A \equiv A$ | (T13) $A \wedge (B \rightarrow A) \equiv A$ |
| (T4)  **false** $\vee A \equiv A$ | (T14) $A \wedge B \rightarrow C \equiv A \rightarrow (B \rightarrow C)$ |
| (T5)  $A \rightarrow$ **true** $\equiv$ **true** | (T15) $A \rightarrow (B \rightarrow C) \equiv B \rightarrow (A \rightarrow C)$ |
| (T6)  $A \rightarrow$ **false** $\equiv \neg A$ | (T16) $(A \rightarrow B) \wedge (A \rightarrow C) \equiv A \rightarrow (B \wedge C)$ |
| (T7)  **true** $\rightarrow A \equiv A$ | (T17) $(A \rightarrow C) \wedge (B \rightarrow C) \equiv (A \vee B) \rightarrow C$ |
| (T8)  **false** $\rightarrow A \equiv$ **true** | (T18) $B \rightarrow (A \rightarrow B)$ |
| (T9)  $A \rightarrow B \equiv \neg A \vee B$ | (T19) $(A \rightarrow C) \rightarrow (A \wedge B \rightarrow C)$ |
| (T10) $A \rightarrow B \equiv \neg B \rightarrow \neg A$ | |
| (T20) $(A \rightarrow C) \wedge (B \rightarrow D) \rightarrow [(A \wedge B) \rightarrow (C \wedge D)]$ | |

*Table 1: Some facts from propositional logic.*

**Definition 2.** (Percolation Pattern) We say that the calculation of effective constraints in subclasses applies the *percolation pattern* if OR-ing (for method preconditions) and AND-ing (for method postconditions and class invariants), respectively, is used along the class hierarchy ([Binder99] and [Binder00]).

**Notation.** (Dispatch mechanism). To denote the common single dispatch mechanism used in object oriented software, expressions like $cl_C.o_{SC}.m(-2)$ will be used: A navigation expression *cl* that is statically typed with class *C* in client code (occasionally, we will use *C-client* as a shorthand) holds a pointer or reference to an instance $o_{SC}$ of class *SC*, and calls method *m* with argument -2. Such a call results in the invocation of $m_{SC}$. When we will speak about an arbitrary, but otherwise fixed, method *m*, we usually will not mention it explicitly and simply write such calls as e.g. $cl_C.o_{SC}$.

**Definition 3.** (Refinement)  Let *S* be a subclass of *T*. And let $spec_S := \langle pre_S, post_S \rangle$, be the specification of an instance method *m* in *S*, and $spec_T := \langle pre_T, post_T \rangle$ its specification in *T*. Then $spec_S$ *refines* $spec_T$, in symbol $spec_S \sqsupseteq spec_T$, if and only if for all calls of *m* where the receiver's dynamic type is a subtype of *S*, every correct implementation of $spec_S$ satisfies $spec_T$. (**Note:** This is an adaptation of Definition 3 in [Leave06]. But be careful when diving into literature about software specification: Usually refinement between the efficient constraints is meant.)

**Definition 4.** (Hierarchy violation) A *hierarchy violation* occurs if a method level constraint for a method in type *T* and the one for an overriding method in a subtype *S* of *T*, or the class invariants of *T* and *S*, have different evaluation results.

**Definition 5.** (Modular reasoning) If it is possible to establish properties of object-oriented code using the static types of expressions (especially of pointers to objects) without need to inspect any of the subclasses involved, then we say that *modular reasoning* can be performed.

**Definition 6.** (Client/view conformance) A class hierarchy has the property of *client conformance*, if a client is never confronted with a surprising or unsafe execution of a server method, and is never hampered by post-condition failures outside its own class scope.

**Notation.** (Percolation pattern variants) In order to allow for concise formulations we introduce the following shorthands for the variants of the percolation pattern that will be important in subsequent considerations in this paper:

- *s-**PRE*** : *simple precondition percolation* (see section 4.1)
- *cl-**PRE*** : *client conforming precondition percolation* (see section 4.2)
- *s-**POST*** : *simple postcondition percolation* (see section 5.1.1)
- *g-**POST*** : *guarded postcondition percolation* (see section 5.1.2)
- *cl-**POST*** : *client conforming postcondition percolation* (see section 5.2)

# 3   Substitutivity: The concept of behavioral subtyping

Late binding of method calls (i.e. binding at run-time which depends on the actual class of the call's receiver) allows flexible code reuse but complicates formal reasoning significantly, as the class of a method call's receiver cannot be statically determined. Usually, object-oriented program systems are designed under an open world assumption: By and by, class hierarchies grow by adding subclasses that may extend its superclasses with new methods, possibly overriding existing ones. Such method redefinitions in subclasses can completely change the semantics of their superclass methods, unless effective mechanisms enforce the preservation of behavioral properties in a suitable way. The dominant solution to this problem is called "behavioral subtyping" for which several approaches have been proposed during the last 25 years.

An important concern in object-oriented software development is how one can reason about consistent extensions of an already existing class hierarchy. In order to preserve modular reasoning it is necessary that each syntactical subtype (subclass) used in a program also is, in a way to be precisely defined soon, a behavioral subtype of each of its supertypes.

## 3.1   Contracts and inheritance

The basic rule governing the relationship between inheritance and assertions is that in a subclass all the superclass contracting assertions (i.e. routine pre- and postconditions, and the class invariants) still apply. Inheritance of assertions guarantees that the behavior of a class is compatible with that of its ancestors: The assertions specify a range of acceptable behaviors for the routine and its eventual redefinitions in subclasses which may specialize this range, but not violate it. In other words, a subtype must not require more or promise less than its supertype(s) ([Meyer97], section 16.1).

## 3.2   The Liskov substitution principle

Modular reasoning is of paramount importance for extensible software systems, in which the set of subclasses of a given class is open. The advantage of modular reasoning is that unchanged methods of client code do not have to be respecified and reverified when new behavioral subtypes are added to class libraries.

*Behavioral subtyping* is a technique for preventing unexpected behavior in a modular way: it ensures that any reasoning that has been done about the behavior of a piece of client code that uses objects of a base class $C$ continues to hold if the code is instead applied to objects of a subclass $SC$ of $C$, i.e. it remains valid if calls to a method $m$ are dispatched to $m_{SC}$ instead of $m_C$. This preservation of reasoning results can be achieved if methods redefined in subclasses satisfy their base class specification, i.e. if certain relationships between the contracting assertions of a subclass and the ones of its superclass(es) hold. This is what the Liskov Substitution Principle (LSP) for the construction of object-oriented software states (see *Figure 1*, a snapshot taken from [Liskov88], section 3.3; see also [LisWin94]).

> A type hierarchy is composed of subtypes and supertypes. The intuitive idea of a *subtype* is one whose objects provide all the behavior of objects of another type (*the supertype*) plus something extra. What is wanted here is something like the following substitution property [6]: If for each object $o_1$ of type S there is an object $o_2$ of type T such that for all programs P defined in terms of T, the behavior of P is unchanged when $o_1$ is substituted for $o_2$, then S is a subtype of T. (See also [2], [17] for other work in this area.)
>
> We are using the words "subtype" and "supertype" here to emphasize that now we are talking about a semantic distinction. By contrast, "subclass" and "superclass" are simply linguistic concepts in programming languages that allow programs to be built in a particular way. They can be used to implement subtypes, but also, as mentioned above, in other ways.

*Figure 1  The presumably first formulation of the LSP*

The basic principle here is as simple as it is important: Subtype instances should not perform any action that will invalidate the assumptions made by clients of a superclass type. From the various approaches that have been proposed for defining behavioral subtyping (see e.g. [Toth05]), for historical and theoretical reasons we will focus on the following two:

*plug-in* ([ZarWin97]):

$$(pre_C \rightarrow pre_{SC}) \wedge (post_{SC} \rightarrow post_C) \quad (1)$$

*relaxed plug-in* ([CheChe00]); also called *Methods Rule* in [Liskov01], p.176):

$$(pre_C \rightarrow pre_{SC}) \wedge (pre_C \wedge post_{SC} \rightarrow post_C) \quad (2)$$

Let us see where implications enter into this scenario: "The notion of expecting less than is guaranteed takes the form of a logical implication: guarantees imply expectations. Likewise, an implication can be used to express that more might be provided than is required: the provided implies the required." (from [Szyp98], p.73).

### 3.3 The key role of the percolation pattern

If we consider specifications as constraints for the range of admissible behavior of methods, then, according to the LSP, an overriding method in a subclass must fit into the range defined by its superclass. This is usually cast in a kind of plug-in condition, as e.g. given in (1) above. An overriding method in subclass $S^kC$ is open for calls of all possible kinds of clients, i.e. of clients from an arbitrary of its superclasses. This imposes the following restrictions on *m*'s effective constraints in a subclass: its effective precondition must not be stronger, and its effective postcondition must not be weaker than anyone given in an overriden method in a superclass. Thus, we get the following result for our example hierarchy (by (T17) and T16)):

$$(pre_C \rightarrow effPre_{SSC}) \wedge (pre_{SC} \rightarrow effPre_{SSC}) \wedge (pre_{SSC} \rightarrow effPre_{SSC}) \equiv (pre_C \vee pre_{SC} \vee pre_{SSC}) \rightarrow effPre_{SSC} \quad (3)$$

$$(effPost_{SSC} \rightarrow post_C) \wedge (effPost_{SSC} \rightarrow post_{SC}) \wedge (effPost_{SSC} \rightarrow post_{SSC}) \equiv effPost_{SSC} \rightarrow (post_C \wedge post_{SC} \wedge post_{SSC}) \quad (4)$$

Thus, the disjunction of preconditions is an upper bound (taking as order: weak ≤ strong; thus $B \le A := A \rightarrow B$), and the conjunction of postconditions is a lower bound for what an overriding method in a subclass must accept and provide, respectively.

Disjunction of preconditions and the conjunction of postconditions thus arise as natural candidates for calculating effective constraints in subclass methods. Moreover, due to (T1) and (T2), they automatically yield the desired plug-in property (e.g. $pre_C \rightarrow pre_C \vee pre_{SC}$ and $post_C \wedge post_{SC} \rightarrow post_C$) independent from the semantic content of the developer written assertions on method level. These facts together with the simplicity of its construction are maybe the reason for the dominant role of the percolation pattern: Eiffel's DbC mechanism builds the effective constraints in subclasses by OR-ing (for method preconditions) and AND-ing (for method postconditions and class invariants), and so do most of the tools, macro packages, and preprocessors for Java, C++, C# etc. we have seen since the mid-nineties. Therefore, it makes sense to have a closer look at percolation in order to gain some knowledge about its properties.

## 4 Calculation of effective preconditions

### 4.1 Traditional precondition percolation and its problems

We call the disjunctive aggregation along a class hierarchy applied in Eiffel's DbC mechanism for building the effective preconditions for overriding methods in subclasses *simple precondition percolation* (*s-**PRE***): it is the most elementary logical formula that makes sense for the purpose of runtime checks. The effective preconditions are constructed as follows (remember $C = S^0C$):

$$effPre_C := pre_C, \text{ and for } k \ge 1 \ effPre_{S^kC} := pre_{S^kC} \vee effPre_{S^{k-1}C}.$$

Thus $effPre_{S^{k-1}C} \rightarrow effPre_{S^kC}$, i.e. an effective superclass precondition implies that of its subclass by (T2) and thus fulfills the plug-in property requested in the first part of (1) in section 3.2.

As mentioned above, disjunctive aggregation of preconditions is the common way to calculate the effective precondition in an overriding method. Let us now consider calls to subclass methods more closely by taking a look at some of the possible configurations shown in *Table 2*. Remember that the effective precondition of e.g. $m_{SSC}$ is given as $effPre_{SSC} := pre_C \vee pre_{SC} \vee pre_{SSC}$, and $effPre_C := pre_C$.

**(P1)** (Surprising execution) Column 1 shows what may be regarded a serious deficiency of *s-PRE*: it does not provide *success conformance* for effective preconditions: *effPre*$_{SSC}$ holds, but *effPre*$_{SC}$ does not. In the example configuration, modular reasoning (section 2.2, Definition 5) for clients with static type *C* or *SC* becomes impossible on the basis of their specification. So, even in the absence of a hierarchy violation (from *pre*$_{SC}$ to *pre*$_{SSC}$) in terms of logical implication (since *pre*$_{SC}$ → *pre*$_{SSC}$ ≡ **true** by (T8)) *s-PRE* can cause irritation for such clients: *cl*$_{SC}$ will be surprised by an execution of *m*$_{SSC}$, whereas *pre*$_{SC}$ ≡ **false** indicates a precondition failure.

**(P2)** (Surprising execution, unsafe execution) Consider the precondition hierarchy violation in column 2 (with *pre*$_{SC}$ → *pre*$_{SSC}$ ≡ **false**). *m*$_{SSC}$ will be executed by the call *cl*$_C$ .*o*$_{SSC}$ although *pre*$_{SSC}$ itself evaluates to **false**, because (due to disjunctive aggregation) *effPre*$_{SSC}$ becomes **true**. This kind of deficiency has first been mentioned in [Karao99], section 4.1. The execution of *m*$_{SSC}$ is both surprising and unsafe. Remarkably, the problem is caused by a class that is not really involved in the client-server relation!

**(P3)** (Unsafe execution) Column 3 also gives another example where a method can be executed although its own specific precondition evaluates to **false.** This is again due to a hierarchy violation on method level, in this case from *pre*$_C$ to *pre*$_{SC}$. But a *C*-client, as expected, would find *m*$_C$ or any of its overriding methods being executed (*m*$_{SC}$ for our example call). If, however, due to dynamic binding *m*$_{SC}$ or *m*$_{SSC}$ would be called, then *cl*$_C$ should be prepared for some surprises: for both of them the effective precondition evaluates to **true**, whereas the method specific one to **false**

|   | 1 | 2 | 3 | 4 |
|---|---|---|---|---|
| *pre*$_C$ | false | false | true | false |
| *pre*$_{SC}$ | false | true | false | false |
| *effPre*$_{SC}$ | false | true | true | false |
| *pre*$_{SSC}$ | true | false | false | false |
| *effPre*$_{SSC}$ | true | true | true | false |
| (P1) | *cl*$_{SC}$ .*o*$_{SSC}$ |  |  |  |
| (P2) |  | *cl*$_C$ .*o*$_{SSC}$ |  |  |
| (P3) |  |  | *cl*$_C$ .*o*$_{SC}$ |  |

*Table 2 Example configurations for s-PRE*

(**Note:** The arrows in columns 2 and 3 denote hierarchy violations in the sense of **false** implications.)

In short, *s-PRE* is too liberal for the following reasons:
1. Clients can be confronted with *surprising executions*: Though the client-specific precondition of a method evaluates to **false,** the effective precondition on server side gives **true**, and the method starts an unexpected execution: (P1), (P2)
2. Clients can be confronted with u*nsafe executions*: The effective precondition on server side gives **true**, but the method specific evaluates to **false**, and the method starts an execution on an insecure basis: (P2), (P3)

### 4.2  Client conformant preconditions

From *Table 2* it is easy to see that the common cause for the problems identfied so far are hierarchy violations which makes the (dynamic) situation of the server different from the (static) expectation of the client. In *Table 3* we propose an alternative, more 'customer oriented' behavior (for a server *o*$_{SC}$ and the possible client views *cl*$_C$ and *cl*$_{SC}$). Each of the columns in the table shows a possible configuration of the method level precondition evaluations and what we suggest as appropriate reactions on the listed calls in dependence of the static type of the caller (client view). Reacting as shown in *Table 3* avoids the problems (P1) - (P3).

We do now know the desirable properties for runtime checking preconditions in subclass methods. As already stated above, an overriding method in a subclass must be open for calls of clients from an arbitrary of its superclasses. Traditionally, this is reflected by using *s-PRE* for constructing the effective preconditions for a subclass method as the disjunction of the method level preconditions. Consequently, a subclass method applies identical checks whenever it is invoked, independent from the client's actual type. This server focused view leads to the problems identified in section 4.1.

|        | 1    | 2     | 3     | 4     |
|--------|------|-------|-------|-------|
| $pre_C$  | **true** | **true**  | **false** | **false** |
| $pre_{SC}$ | **true** | **false** | **true**  | **false** |
| $cl_C.o_{SC}$ | ✓ | ⊠ | ⊗ | ✗ |
| $cl_{SC}.o_{SC}$ | ✓ | ⊗ ⊠ | ✓ | ✗ |

*Table 3: Client conforming preconditions*

✓ ... accept call (same as for *s-**PRE***)
✗ ... reject call (same as for *s-**PRE***)
⊗ ... reject call (contrary to *s-**PRE***: surprising execution)
⊠ ... reject call (contrary to *s-**PRE***: unsafe execution)

In *Table 3* it is shown that taking care of the client view can avoid these problems. In the following we will use *client_T* as a shorthand for "client has static type T", $view_T := client_T \wedge pre_T$ to denote the client's view on a method *m*, and e.g. *effView_{SSC}* := $view_C \vee view_{SC} \vee view_{SSC}$. The task to be done is symmetrical: We must let subclass methods $m_S$ combine the views of its possible clients with their own safeguard for execution, and we must enable clients to enforce their view on the program state upon its possible servers. A nearby approach is to define

$$effView_S := \bigvee\nolimits_{C \in supers(S)} view_C \quad \text{and} \quad effConPre_S := effView_S \wedge pre_S. \tag{5}$$

The *effective client conforming precondition* of subclass method $m_S$, *effConPre_S*, is thus open for all admissible client views and additionally checks for safe execution. Expanding (5) we get

$$effConPre_S := \bigvee\nolimits_{C \in supers(S)} (view_C \wedge pre_S) \tag{6}$$

(6) is called *client conforming precondition percolation* (*cl-**PRE***). Illustrated by our standard example hierarchy, this gives

$$effConPre_{SSC} := (view_C \wedge pre_{SSC}) \vee (view_{SC} \wedge pre_{SSC}) \vee (view_{SSC} \wedge pre_{SSC}).$$

It is clear that all but one of the client predicates become **false**; hence all the corresponding conjunctions evaluate to **false** and therefore act as a neutral contribution to the overall evaluation of *effConPre_{SSC}*. Therefore, this evaluation result is determined by the conjunction corresponding to the static type of the client: If both the client's and also the server's precondition evaluate to **true**, then also *effConPre_{SSC}* as a whole. This gives the desired behavior indicated in *Table 3*.

## 5 Calculation of effective postconditions and class invariants

### 5.1 Traditional postcondition percolation and its problems

#### 5.1.1 Eiffel's approach

Analogous to preconditions, using $\wedge$ instead of $\vee$, the effective postconditions (and also class invariants) are constructed for subclasses, e.g. as

$$effPost_C := post_C, \text{ and for } k \geq 1 \quad effPost_{S^kC} := post_{S^kC} \wedge effPost_{S^{k-1}C}.$$

Thus $effPost_{S^kC} \rightarrow effPost_{S^{k-1}C}$ i.e., an effective subclass postcondition implies that of its superclass, by (T1). We call the conjunctive aggregation along a class hierarchy applied in Eiffel's DbC mechanism for building the effective postconditions for overriding methods in subclasses *simple postcondition percolation* (*s-**POST***). It is the most elementary logical formula that makes sense for the purpose of runtime checks. (*s-**POST***) fulfills the plug-in match at the level of effective constraints by construction; see section 3.2, (1).

As can be seen below, we can easily find problematic situations in *Table 4*, which are similar to those we have discussed for preconditions.

**(P4)** (Surprising failure, in superclass constraint) In column 1 $post_{SC} \equiv$ **true**, whence one would expect a success from a call $cl_{SC}.o_{SC}$. Nevertheless, execution of $m_{SC}$ would be reported to have produced a wrong result, since $effPost_{SC} \equiv$ **false**. This is obviously due to the postcondition hierarchy violation $post_{SC} \rightarrow post_C \equiv$ **false** on method level. You may argue that this is not really a surprise, since a subclass method has to respect superclass constraints. But why - from its point of view - should $cl_{SC}$ be interested in the superclass postcondition of $m_C$?

**(P5)**  (Perfect execution) Column 2 demonstrates another inadequate property of *s-POST*: A subclass method $m_S$, when executed, does not know about the static type (which is always a superclass of *S*) of its current client. Therefore, to be on the safe side, $m_S$ has to give an answer that satisfies all possible clients (by making all superclass postconditions valid). As a consequence methods must deliver the best possible answer even to the least possible request: As soon as only one arbitrary method level precondition is true somewhere in the method's superclass hierarchy, *s-POST* has to fulfill the postconditions of all ancestor methods. But a *C*-client would not bother about *SC*-specific postconditions (or class invariants). Our example call $cl_C.o_{SSC}$ in column 2 would be expected to be successful (since a *C*-client reasons on basis of $post_C$), whereas, due to $post_{SC} \equiv$ **false**, also $effPost_{SSC} \equiv$ **false** (although the execution of $m_{SSC}$ delivers correct results) and a surprising failure would be reported, caused by a method belonging to a class that is not involved at all.

|                  | 1                    | 2                    | 3                                      | 4     |
|------------------|----------------------|----------------------|----------------------------------------|-------|
| $post_C$         | **false**            | **true**             | **true**                               | **true** |
| $post_{SC}$      | **true**             | **false**            | **true**                               | **true** |
| $effPost_{SC}$   | **false**            | **false**            | **true**                               | **true** |
| $post_{SSC}$     | **true**             | **true**             | **false**                              | **true** |
| $effPost_{SSC}$  | **false**            | **false**            | **false**                              | **true** |
| (P4)             | $cl_{SC}.o_{SC}$     |                      |                                        |       |
| (P5)             |                      | $cl_C.o_{SSC}$       |                                        |       |
| (P6)             |                      |                      | $cl_C.o_{SSC}$<br>$cl_{SC}.o_{SSC}$    |       |

*Table 4: Example configurations for s-POST*

(**Note:** The arrows in column 1 and 2 denote hierarchy violations in the sense of a **false** implication.)

**(P6)**  (Surprising failure, in subclass constraint) Let us look in detail what (P6) means for clients: For both calls $cl_C.o_{SSC}$ and $cl_{SC}.o_{SSC}$ $effPost_{SSC}$ is claculated in the same way, since it does not depend on the client's view but only on the receiver's dynamic type. Clients $cl_C$ and $cl_{SC}$, reasoning on their respective class specific postconditions $post_C$ and $post_{SC}$, would be disappointed in their expectation for a successful execution of $m_{SSC}$ (column 3).

In a nutshell, *s-POST* is too strong for the following reasons:
1. Clients can be confronted with *surprising failures*: Though the client-specific postcondition of a method evaluates to **true,** the effective postcondition on server side gives **false**, and execution ends with an unexpected message that the calculated results are invalid: (P4), (P6).
2. Whenever a method is executed it has to perform a *perfect execution*: The simple conjunctive form of effective postconditions for subclass methods says that its own and also all the promises made by its ancestor methods have to be guaranteed after method execution: (P5).

### 5.1.2   Join composition

The just described requirement for perfect executions of subclass methods can be somewhat relaxed if we use guarded postconditions as follows:

$$\text{g-}effPost_{SSC} := (pre_C \rightarrow post_C) \land (pre_{SC} \rightarrow post_{SC}) \land (pre_{SSC} \rightarrow post_{SSC}). \tag{7}$$

We call this pattern the *guarded postcondition percolation* and denote it with *g-POST*. Using this kind of effective constraint, only those postconditions have to be fulfilled for which the preconditions have been true – both at method level. Since implications with **false** antecedent evaluate to **true**, we can take advantage of this syntactical property and have found a simply structured predicate as solution for problem (P5) identified in section 5.1.1. Reformulating the above, we can also say that all postconditions have to be fulfilled for which the preconditions have been true – even if they are from other than the client's class. But from a client's point of view this is an unecessarily strong requirement.

What we have just presented is the way how contracts are built in case of so-called *join composition*. Using guarded postconditions for a method's specification means to partition the range of legal states for its execution.

This mechanism can thus be used for software construction out of parts: Assume you have a set of possible clients, each with its own specific expectations for certain method: if $P_i$ at start, then $Q_i$ after computation, written as (method) specification $\langle P_i, Q_i \rangle$. These requirements can consistently be combined by join composition in the following way: $\langle \textit{pre-join}, \textit{post-join} \rangle := \langle \vee_i P_i, \wedge_i (P_i \rightarrow Q_i) \rangle$ is the weakest specification that refines each of $\langle P_i, Q_i \rangle$, $1 \leq i \leq n$; for more details see the *Appendix*, **2**. Obviously, this is again an instantiation of the percolation pattern, whence the plug-in property is preserved for this variant of calculating the effective constraints. This kind of contract has also been used in slightly different or more formal contexts also in [DSouza99], section 8.2.1; [HeKBau04], section 4.1; [LeiMan99], section 5; and [Dunne02], section 3.3

(**P7**) (Surprising failure, , in subclass constraint) Though guarded postconditions do help with respect to some aspects, we are still not at the end of our search, because they do not fully provide what we are really looking for: Consider a call $cl_C.o_{SC}$ and assume that $pre_C$, $pre_{SC}$ and $post_C$ evaluate to **true**, and $post_{SC}$ to **false**; then a safe execution has been enabled and yet, by g-$\textit{effPost}_{SC} \equiv (pre_{SC} \rightarrow post_{SC}) \equiv$ **false**, $cl_C$ gets confronted with an error for a state that it is not really concerned with from its point of view.

Since effective preconditions follow *s-**PRE***, we can limit further discussion to how the postcondition problems of section 5.1.1 behave under join composition:

- In case of (P4), a simple calculation shows that the effective postcondition for the call $cl_{SC}.o_{SC}$ becomes completely independent of $m_{SC}$'s own specification: g-$\textit{effPost}_{SC} := (pre_C \rightarrow post_C) \wedge (pre_{SC} \rightarrow post_{SC}) \equiv (pre_C \rightarrow$ **false**$) \wedge (pre_{SC} \rightarrow$ **true**$) \equiv (pre_C \rightarrow$ **false**$) \equiv \neg pre_C$. It is quite astonishing that the value of the effective postcondition is determined by the negation of a precondition ($A \rightarrow$ **false** $\equiv \neg A$ is a law of propositional logic) and, moreover, by a class not involved at all into the call $cl_{SC}.o_{SC}$: If $pre_C$ fails, then $cl_{SC}$ will be served as expected, otherwise a surprising failure occurs.
- Concerning (P5) join composition partially helps: Only the postconditions for those classes are requested to be valid, for which the preconditions have already been. But, due to surprising executions in *s-**PRE***, other clients can also become possible callers. As example take a call $cl_C.o_{SC}$ and the configuration of column 2: we get (using similar calculations as above) g-$\textit{effPost}_{SC} \equiv \neg pre_{SC}$. This time g-$\textit{effPost}_{SC}$ depends on method's $m_{SC}$ own specification, but again in a rather astonishing way: Only if its precondition becomes **false,** then its effective postcondition evaluates to **true**.
- Finally we reconsider (P6): For the calls $cl_C.o_{SSC}$ and $cl_{SC}.o_{SSC}$ in column 3 of *Table 4* we get: g-$\textit{effPost}_{SSC} := (pre_C \rightarrow$ **true**$) \wedge (pre_{SC} \rightarrow$ **true**$) \wedge (pre_{SSC} \rightarrow$ **false**$) \equiv \neg pre_{SSC}$. We see that both calls get the same answer which is not necessarily adequate. Besides, why should clients $cl_C$ or $cl_{SC}$ get involved into subclass constraints at all? For both of them $m_{SSC}$'s execution has done a perfect job from their own point of view.

  More concretely, we know that at least one of the method preconditions in the triple ($pre_C$, $pre_{SC}$, $pre_{SSC}$) has been **true**. Assume we had (**false**, **false**, **true**); then g-$\textit{effPost}_{SSC} \equiv$ **false** and we had a safe execution ($pre_{SSC} \equiv$ **true**) that is surprising, but fits for both clients $cl_C$ and $cl_{SC}$; but a failure is reported. As a second example for preconditions assume (**true**, **false**, **false**); then g-$\textit{effPost}_{SSC} \equiv$ **true**. This time we had an execution as expected from $cl_C$ but surprising for $cl_{SC}$, and it was unsafe. So there may well be reasons to ask if we could do better than that.

Summarizing the results of section 5.1 we can say: Both *s-**POST*** and join composition are too strict (although the latter is weaker than the former) since they disable some results from being accepted by clients.

## 5.2 Client conforming postcondition percolation

The postcondition handling suffers from reasons analogous to that which have been described for preconditions in section 4.1: *Table 4* indicates that it is again the different evaluations in the (dynamic) situation of the server and the (static) expectation of the client that are the crucial point.

Reacting as shown in *Table 5* avoids the problems (P4) -(P6):

- In (P4) we had $post_{SC} \equiv$ **true** and $post_C \equiv$ **false**, whence *s-**POST*** gives $\textit{effPost}_{SC} := post_C \wedge post_{SC} \equiv$ **false**, and the execution of method $m_{SC}$ by a call $cl_{SC}.o_{SC}$ would be reported to have produced a wrong result although an $cl_{SC}$ client clearly would expect a success. The entry in line 2 and column 3 accepts the result of this execution.
- For (P5) we get the relaxation as desired: Only the postcondition corresponding to client's view must be satisfied.
- Finally (P6): For the calls $cl_C.o_{SSC}$ and $cl_{SC}.o_{SSC}$ and using the configuration of column 3 in *Table 4* we get $\textit{effPost}_{SSC} := post_C \wedge post_{SC} \wedge post_{SSC} \equiv$ **false** using s-**POST**; but according to *Table 5* we would check the clients postcondition only and accept the results, and that fits the expectation of both $cl_C$ and $cl_{SC}$.

|  | 1 | 2 | 3 | 4 |
|---|---|---|---|---|
| $post_C$ | **true** | **true** | **false** | **false** |
| $post_{SC}$ | **true** | **false** | **true** | **false** |
| $cl_C.o_{SC}$ | ✓ | ☑ | ✗ | ✗ |
| $cl_{SC}.o_{SC}$ | ✓ | ✗ | √ | ✗ |

*Table 5: Client conforming postconditions*

✓ ... accept result (same as for *s-POST*)
✗ ... reject result (same as for *s-POST*)
☑ ... accept result (contrary to *s-POST*: subclass state failure)
√ ... accept result (contrary to *s-POST*: superclass state failure)

Analogous to the precondition situation of section 4.2 we have found the desirable properties for runtime checking postconditions in subclass methods.

Using a simple example configuration, let us now examine if *cl-PRE* together with *g-POST*, the less restrictive of the traditional postcondition aggregations, works as desired: e.g. *g-effPost$_{SC}$* := $(pre_C \rightarrow post_C) \land (pre_{SC} \rightarrow post_{SC})$. Due to combinatorial reasons, the evaluations of pre- and postconditions can vary independently from each other. And because all possible configurations are of interest for our analysis, the representation becomes a little bit more complex than that in *Table 5*. Tedious, but straight-forward calculations lead to the results shown in *Table 6*, where client conforming precondition evaluation according to *Table 3* has been applied.

The shaded part presents plausible client behavior, a kind of confidence requirement: As soon as a method's calculations have been done, the only relevant thing is, if the result matches the client's request. If the postcondition visible to the client (i.e. the one of its static type that is used for reasoning) evaluates to **true**, then the client accepts the result, even if a subclass method's own postcondition fails. Thus we leave the detection of possible errors caused by the failure of a subclass postcondition for a later occasion: perhaps we are lucky, and things run fine for the current client.

|  | 1 | 2 | 3 | 4 |  |  |  |  |  |
|---|---|---|---|---|---|---|---|---|---|
| $post_C$ | **true** | **true** | **false** | **false** | | | | | |
| $post_{SC}$ | **true** | **false** | **true** | **false** | | | | | |
| $cl_C.o_{SC}$ | ✓ | ✓ | ✗ | ✗ | | | | | |
| $cl_{SC}.o_{SC}$ | ✓ | ✗ | ✓ | ✗ | 1 | 2 | 3 | 4 | |
|  | $(pre_C \rightarrow post_C) \land (pre_{SC} \rightarrow post_{SC})$ | | | | $(pre_C \rightarrow post_C, pre_{SC} \rightarrow post_{SC})$ | | | | $(pre_C, pre_{SC})$ |
| $cl_C.o_{SC}$ / $cl_{SC}.o_{SC}$ | ✓ / ✗ | ☑ / ✗ | ✗ / ☑ | ✗ / ✗ | (t, t) | (t, f) | (f, t) | (f, f) | (true, true) |
| $cl_C.o_{SC}$ / $cl_{SC}.o_{SC}$ | - | - | - | - | - | - | - | - | (true, false) |
| $cl_C.o_{SC}$ / $cl_{SC}.o_{SC}$ | - / ✓ | - / ✗ | - / ✓ | - / ✗ | (t, t) | (t, f) | (t, t) | (t, f) | (false, true) |
| $cl_C.o_{SC}$ / $cl_{SC}.o_{SC}$ | - | - | - | - | - | - | - | - | (false, false) |

*Table 6: cl-PRE combined with g-POST*

✓ ... accept result (same as for *g-POST*)
✗ ... reject result (same as for *g-POST*)
☑ ... accept result (contrary to *g-POST*)
- ... call rejected (due to *Table 3*), hence no postcondition check

The lower half is made up of two parts: the right one holds the results for the 16 possible configurations of constraint evaluations represented as pairs of implication truth values, and the left one gives the overall evaluation for *g-effPost$_{SC}$* for these 16 cases. In line 1 of the lower part, which is relevant for both *cl$_C$* and *cl$_{SC}$* clients (since both *pre$_C$* and *pre$_{SC}$* are **true**), we see that *g-effPost$_{SC}$* would evaluate to **false** if one of the postconditions is **false:** for the call *cl$_C$.o$_{SC}$* we would like *pre$_{SC}$* → *post$_{SC}$* ≡ **false** to be neglected in column 2, and for *cl$_{SC}$.o$_{SC}$* *pre$_C$* → *post$_C$* ≡ **false** in column 3, since both failures are outside the client's view. Hence some special checks are in order to achieve the desired behavior, marked as ☑.

We define client conforming effective postcondition as follows:

$$\textit{effConPost}_S := \bigwedge\nolimits_{C \in \textit{supers}(S)} \textit{conPost}_C, \text{ where } \textit{conPost}_C := (\textit{view}_C \rightarrow \textit{post}_C). \tag{8}$$

It is clear that all but one of the client predicates become **false**; hence all the corresponding implications evaluate to **true** and therefore act as a neutral contribution to the overall evaluation of *effConPost$_S$*. This evaluation result is determined by the implication corresponding to the static type of the client: If the client's postcondition evaluates to **true**, then also *effConPost$_S$* as a whole. This gives the desired behavior indicated in **Fehler! Verweisquelle konnte nicht gefunden werden.**. (8) is called the *client conforming postcondition percolation* (*cl-**POST***). Let us see how client conforming effective postconditions work for the problems analysed above in section 5.1.1.

- It clearly solves (P4): *effConPost$_{SC}$* calculates to (*view$_C$* → *post$_C$*) ∧ (*view$_{SC}$* → *post$_{SC}$*) ≡ (*view$_C$* → **false**) ∧ (*view$_{SC}$* → **true**) ≡ (*view$_C$* → **false**) ≡ ¬(*client$_C$* ∧ *pre$_C$*) ≡ **true**, since our call was *cl$_{SC}$.o$_{SC}$*.
- For (P5) we get a further relaxation compared to guarded postconditions: Only the postcondition corresponding to the static type of the calling client must be satisfied.
- Finally (P6): For the calls *cl$_C$.o$_{SSC}$* and *cl$_{SC}$.o$_{SSC}$* and using the configuration of column 3 in *Table 4* we get: *effConPost$_{SSC}$* := (*view$_C$* → **true**) ∧ (*view$_{SC}$* → **true**) ∧ (*view$_{SSC}$* → **false**) ≡ ¬*view$_{SSC}$* ≡ **true**, and that fits the expectation of both *cl$_C$* and *cl$_{SC}$*. A similar argument applies to (P7) at the end of section 5.1.2.

## 5.3 Client conformance for class invariants

Class invariants are the last aspect we will consider in our analysis of behavioral subtyping and substitutivity. For both simple percolation and specification inheritance the calculation of effective invariants in subclasses follows Eiffel's Design by Contract™ paradigm for postconditions, i.e. conjunctive aggregation of the class specific invariants is used. So we get e.g. *effInv$_{SSC}$* := *inv$_C$* ∧ *inv$_{SC}$* ∧ *inv$_{SSC}$*, and thus also the same problems that we have already analysed in section 5.1.

It is argued that "Class invariants simplify the specification of methods by factoring out common properties" ([HuiKui00], section 3), or that an "*object invariant* (sometimes incorrectly called class invariant) is a property on the object fields that holds in the steady states of the object, i.e., it is at the same time a precondition and a postcondition of all the methods of a class." ([BouLog12], Introduction). Since this view on invariants can hardly be abandoned, what should we have to conclude? Something like an "invariance principle for invariants": as a precondition, an invariant must not be strengthened, and as a postcondition not be weakened. This is counterintuitive, since we do, of course, want to formulate additional properties for subclass objects with additional fields. So we are again confronted with the extended state space problem.

But handling invariants in the same way as we did with preconditions, we can avoid this problem. So we define (see (5) of section 4.2)

$$\textit{effConInvPre}_S := \bigvee\nolimits_{C \in \textit{supers}(S)} (\textit{client}_C \wedge \textit{inv}_C) \wedge \textit{inv}_S, \tag{9}$$

where we have again a safety check for the server method: We want to make sure that *m$_S$* can start its work with an instance in a valid state. Doing the analogue for postcondition gives (see (8) of section 5.2)

$$\textit{effConInvPost}_S := \bigwedge\nolimits_{C \in \textit{supers}(S)} (\textit{client}_C \wedge \textbf{old\_inv}_C \rightarrow \textit{inv}_C), \tag{10}$$

where **old_inv$_C$** denotes the evaluation of the invariant before server method execution, i.e. in the program state of precondition evaluation. We will come back to invariants in section 6.3 below.

# 6 Client conformance in class hierarchies

## 6.1 Summary and analysis

Traditionally calculated effective constraints are server focused in the following sense: for e.g. method *m$_{SCC}$*, there are three kinds of possible clients: *cl$_C$*, *cl$_{SC}$*, and *cl$_{SCC}$*. But the effective precondition is the same for each of them, because the client's static type is unknown to *m$_{SCC}$*, which usually has the task to check the constraints for its clients. In order to be able to classify $m_{SCC}$ as having executed correctly in all cases, the method must satisfy the expectation of each possible kind of clients. This is an nonsymmetric strategy enforced on servers: Called by an arbitrary one of their clients, they must be perfect and have to serve all of them. As a result the commonly used construction mechanisms for effective constraints depend on the server only and completely neglect the client.

It has already been said that the type of the client (i.e. of some suitable navigation expression in the chosen implementation language) is known at compile time (static), whereas the type of the server is not (dynamic). This is an external viewpoint that induces an implicit asymmetry between client and server. But when regarded from the inside we get a complete symmetry: There are the two groups of clients and servers which have to organize their cooperation in a suitable way: For each client type there is a set of possible servers (for e.g. $cl_C$ : $m_C$, $m_{SC}$ and $m_{SCC}$), and for each server there is a set of possible clients (for e.g. $m_{SCC}$ : $cl_C$, $cl_{SC}$, and $cl_{SCC}$); and neither clients nor servers do know which partner they will have to work with during execution. Therefore, both of them must be more careful in the formulation of the cooperation contract, which has to take into account all possible configurations appropriately.

Our considerations have directed attention to two groups of concepts that we have called *client view*, and *surprising* and *unsafe execution* of methods, or *surprising failures*. This suggests the idea that preconditions may be ascribed a double role in method specifications: On the one hand they serve to describe the view a client has on a certain situation, and thus also the information it uses for reasoning about a program state. On the other hand they express what a server method requires for starting its work and for delivering correct results. These two aspects have to be mirrored in effective constraints in a way that must both protect a client's view and also provide the necessary working conditions for the server.

The more elaborate negotiation of the conditions that a client accepts and under which a server is willing to do its work in client conforming method specifications finally results in a fair modification of the classical approaches to behavioral subtyping: Servers are allowed to refuse work in some additional situations (see *Table 3*), but – in exchange for this – they must make their results acceptable for clients more often (see *Table 5* and *Table 6,* and Proposition 2 below).

## 6.2 Logical properties of client conformant method specifications

In this section with will take a short look on the logical properties of client conforming specifications (assuming $S$ to be a subtype of $T$):

**Lemma.** $effView_T \rightarrow effView_S$

Proof. By (T2) we have $effView_T \rightarrow effView_T \vee effView_S$, and by definition (see (5) in section 4.2), $effView_T \vee effView_S \equiv effView_S$. Therefore, $effView_T \rightarrow effView_S$ , by (T11).

It is now easy to show that a subclass specification refines that of a superclass, given that a safe execution of the subclass server method, say $m_S$, can be done (i.e. $pre_S$ holds) for a client of type $T$ (i.e. $view_T$ holds):

**Proposition 1.** $view_T \wedge pre_S \rightarrow \langle effConPre_S, effConPost_S\rangle \sqsupseteq \langle effConPre_T, post_T\rangle$

Proof. (1) From $pre_S$ we get $pre_T \rightarrow pre_S$ by (T18). Using the above Lemma and applying (T20) to ($effView_T \rightarrow effView_S$) $\wedge$ ($pre_T \rightarrow pre_S$) gives ($effView_T \wedge pre_T$) $\rightarrow$ ($effView_S \wedge pre_S$), i.e. $effConPre_T \rightarrow effConPre_S$.

(2) Assume $effConPost_S$ holds; then by (T1), $effConPost_S \rightarrow conPost_T$, and applying modus ponens we get $conPost_T \coloneqq (view_T \rightarrow post_T)$. Assuming that $view_T$ holds (at start of $m_S$) one immediately gets $post_T$ by modus ponens.

In section 4.1 we concluded with the statement that *s-PRE* is too liberal since it allows surprising and unsafe execution of server methods, and the last sentence of section 5.1 was: "Postcondition percolation as well as join composition are both too strict (although the latter is weaker than the former) since they disable some results from being accepted by clients." To avoid unnecessary formalism we will now use our standard hierarchy examples to show the following

**Proposition 2.** (1) *cl-PRE* is stronger than s-*PRE*, and (2) *cl-POST* is weaker than both *g-POST* and *s-POST*:

Proof. (1) $effConPre_{SSC}$ $\equiv (view_C \wedge pre_{SSC}) \vee (view_{SC} \wedge pre_{SSC}) \vee (view_{SSC} \wedge pre_{SSC})$
$\equiv effView_{SSC} \wedge pre_{SSC}$
$\rightarrow pre_{SSC}$
$\rightarrow pre_C \vee pre_{SC} \vee pre_{SSC}$
$\equiv effPre_{SSC}$

(2) $effPost_{SC} \rightarrow g\text{-}effPost_{SC} \rightarrow effConPost_{SC}$ : Applying modus ponens to $effPost_{SC} \rightarrow post_C$ we get $post_C$. As an instance of (T18) we get $post_C \rightarrow (pre_C \rightarrow post_C)$     (a)
    $\rightarrow pre_C \rightarrow post_C$, by modus ponens
    $\rightarrow client_C \wedge pre_C \rightarrow post_C$, by (T19)     (b)
    $\equiv view_C \rightarrow post_C$, by definition (see section 4.2).

Applying (T20) to (a) and (b), respectively, we finally arrive at the desired results:
$$post_C \wedge post_{SC} \rightarrow [(pre_C \rightarrow post_C) \wedge (pre_{SC} \rightarrow post_{SC})] \quad \text{(a')}$$
$$\rightarrow [(view_C \rightarrow post_C) \wedge (view_{SC} \rightarrow post_{SC})]. \quad \text{(b')}$$

## 6.3 Client conforming RAC

Putting together from sections 4.2 and 5.3 the constraints we want to hold before invoking a server method for calculation we get *effConPre$_S$* ∧ *effConInvPre$_S$*. A short calculation shows that this amounts to *cl-**PRE*** expanded with an invariant check:

$$\textit{effConPre}_S \wedge \textit{effConInvPre}_S \equiv \bigvee\nolimits_{C \in \textit{supers}(S)} (client_C \wedge pre_C) \wedge \bigvee\nolimits_{C \in \textit{supers}(S)} (client_C \wedge inv_C) \wedge pre_S \wedge inv_S.$$

Reconfiguring the disjunctions using distributivity, $A \wedge (B \vee C) \equiv (A \wedge B) \vee (A \wedge C)$, in an example with two classes yields

$[(client_C \wedge pre_C) \vee (client_T \wedge pre_T)] \wedge [(client_C \wedge inv_C) \vee (client_T \wedge inv_T)]$

$\equiv [[(client_C \wedge pre_C) \vee (client_T \wedge pre_T)] \wedge (client_C \wedge inv_C)] \vee [[(client_C \wedge pre_C) \vee client_T \wedge pre_T)] \wedge (client_T \wedge inv_T)]$

$\equiv (client_C \wedge pre_C \wedge inv_C) \vee (client_C \wedge client_T \wedge ...) \vee (client_C \wedge client_T \wedge ...) \vee (client_T \wedge pre_T \wedge inv_T)$

$\equiv (client_C \wedge pre_C \wedge inv_C) \vee (client_T \wedge pre_T \wedge inv_T)],$

since the conjunctions of the form $client_C \wedge client_T$ always evaluate to **false.** So we can define the constraint to be checked before starting server method $m_S$ as a straightforward extension of *effConPre$_S$*:

$$\textit{ConRACPre}_S := [\bigvee\nolimits_{C \in \textit{supers}(S)} (client_C \wedge pre_C \wedge inv_C)] \wedge (pre_S \wedge inv_S)$$

$$\equiv [\bigvee\nolimits_{C \in \textit{supers}(S)} (client_C \wedge pre_C \wedge inv_C \wedge pre_S \wedge inv_S)],$$

and call it the *client conforming runtime pre-check*. As can be seen, $inv_C$ becomes masked in the same way as $pre_C$, and, finally, we have got another variant of the percolation pattern. A *C*-client will thus reason based on the constraint $pre_C \wedge inv_C$, just as one would expect (see section 4.2); in addition to that we want to provide a safe execution of the server method by requiring both $pre_S$ and $inv_S$ to be valid.

Turning to postconditions, from sections 5.2 and 5.3 we get *effConPost$_S$* ∧ *effConInvPost$_S$* as constraints we require to be valid after execution of server method $m_S$:

$$\textit{effConPost}_S \wedge \textit{effConInvPost}_S \equiv \bigwedge\nolimits_{C \in \textit{supers}(S)} (client_C \wedge pre_C \rightarrow post_C) \wedge \bigwedge\nolimits_{C \in \textit{supers}(S)} (client_C \wedge \textbf{old\_}inv_C \rightarrow inv_C).$$

Putting together the constraints for one of the possible clients of server method $m_S$, e.g. $cl_C$, we get

$(client_C \wedge pre_C \rightarrow post_C) \wedge (client_C \wedge \textbf{old\_}inv_C \rightarrow inv_C)$

$\equiv client_C \rightarrow [(pre_C \rightarrow post_C) \wedge (\textbf{old\_}inv_C \rightarrow inv_C)]$, by (T14) and (T16)

$\equiv client_C \rightarrow [pre_C \wedge \textbf{old\_}inv_C \rightarrow post_C \wedge inv_C]$, by (T20)

$\equiv client_C \wedge pre_C \wedge \textbf{old\_}inv_C \rightarrow post_C \wedge inv_C$, by (T14).

So we define the constraint to be checked after server method mS has finished its calculations as

$$\textit{ConRACPost}_S := \bigwedge\nolimits_{C \in \textit{supers}(S)} (client_C \wedge pre_C \wedge \textbf{old\_}inv_C \rightarrow post_C \wedge inv_C)$$

and call it the *client conforming runtime post-check*. Assume that we deal with a *C*-client whose call is dynamically dispatched to the server method $m_S$ in subclass $S$ of $C$. Thus, before invocation of $m_S$ both $pre_C$ and $inv_C$ have been valid. If both the client's postcondition and class invariant evaluate to **true**, then also *ConRACPost$_S$* as a whole. As can be seen, $inv_C$ becomes masked in the same way as $post_C$, and, finally, we again can use a variant of the percolation pattern.

Note that we do not require that the server method's postcondition or the server class invariant hold (unless the client and server classes coincide). Both are unnecessary constraints in a client aware checking policy; it is left to the next server method invoked to check if its class specific precondition and class invariant provide a safe execution, and this server method need not be of the same class $S$.

# 7 Some further aspects of behavioral subtyping

## 7.1 Discussing behavioral subtyping on method level

Up to now we have learned that the prevailing technique for RAC is the percolation pattern, and that one of the reasons for this dominance may well be expressed by the heading of Theorem 3 in [Leave06]: "Specification inheritance forces behavioral subtyping". And we know that this syntactic automatism is due to the laws (T1) and (T2) from propositional logic. In other words, the semantic content of the individual method level constituents has become secondary. On the other hand, in section 3.3 and in the Appendix it is shown that there are plausible justifications for this kind of evaluating constraints over class hierarchies.

But there is yet another view on behavioral subtyping with a stronger focus towards the developer written contracts in form of pre- and postconditions, and invariants. It can be found in the presentation and discussion of many examples and, in a prominent form, in the so-called Methods Rule given in[Liskov01], pp. 176/177. (See also (2) at the end of section 3.2; the emphasizing by *italics* is mine.):

> "(The Methods Rule): A subtype method can weaken the precondition and can strengthen the post-condition.
> Precondition Rule: $pre_{super} \rightarrow pre_{sub}$
> Postcondition Rule: $pre_{super}\ \&\&\ post_{sub} \rightarrow post_{super}$
> Both conditions must be satisfied to achieve compatibility between the sub- and supertype methods.
> Weakening the precondition means that the subtype method requires less from its caller than the supertype method does. This rule makes sense because when code is written in terms of the supertype specification, it must satisfy the supertype method's precondition. *Since this precondition implies that of the subtype, we can be sure that the call to the subtype method will be legal if the call to the supertype method is legal*.
> … the postcondition rule. … This rule makes sense because the calling code depends on the postcondition of the supertype method, but this follows from the postcondition of the subtype method. However, the calling code depends on the method's postcondition only if the call satisfies the precondition (since otherwise the method can do anything);"

One kind of the running examples often found comes from an economical branch that has become so very popular throughout the world during the last few years (taken from [DovJohn10] together with the authors' explanation):

> "**Example 1**. Consider the following two classes:
>
> ```
> class Account {
>     int bal;
>     void deposit(nat x) {update(x)}
>     void withdraw(nat x) {update(-x)}
>     void update(int x) {bal := bal + x}
> }
> class FeeAccount extends Account {
>     int fee;
>     void withdraw(nat x) {update(-(x+fee))}
> }
> ```
>
> In this example, class *Account* implements ideal bank accounts for which the withdraw method satisfies the pre- and postcondition pair ($bal = bal_0$, $bal = bal_0 - x$), where $bal_0$ is a logical variable used to capture the initial value of *bal*. The subclass *FeeAccount* redefines the withdraw method, charging an additional fee for each withdrawal. Thus, class *FeeAccount* is not a *behavioral subtype* of class Account. However, the example illustrates that it might be fruitful to implement *FeeAccount* as an extension of *Account* since much of the existing code can be *reused* by the subclass. In this paper we focus on incremental reasoning in this setting: Subclasses may reuse and override superclass code in a flexible manner such that superclass specifications need not be respected."

If you have a closer look at the situations described in (P1) to (P7) you can easily see that their common characteristic is the same as in the account example above: It is always a hierarchy violation in the sense that the evaluation of a method level constraint differs between a super and subclass somewhere on the way from the server up to the base class. (**Note:** Not always would such a hierachy violation give **false** for a corresponding implication. But that is not a real problem for current RAC practice, because none of the prominent tools does check constraint hierarchies according to the Methods Rule. Instead, by using percolation, something like a Methods Rule on the level of effective constraints is built.)

### 7.1.1   Precondition checking

Let us interpret the Methods Rule (MR) verbatim to see what such explicit hierarchy checks would result in. To keep things simple, assume that we deal with a *C*-client whose call is dynamically dispatched to a method $m_{SSC}$, and that the client precondition $pre_C$ holds. Then the following check should be done if applying MR in a strong way:

$$(pre_C \rightarrow pre_{SC}) \wedge (pre_{SC} \rightarrow pre_{SSC}) \wedge pre_{SSC} \tag{a}$$

Given that (a) is valid as a whole, then repeated application of (T12) yields $pre_C \wedge pre_{SC} \wedge pre_{SSC}$, which certainly is not a weaker constraint than $pre_C$. And concerning properties of an extended state space in subclasses, one will usually fail to prove e.g. ($pre_C \wedge addedPre_{SC}$) from $pre_C$, unless $addedPre_{SC}$ is a tautological truth. No wonder, that current RAC machines do not perform hierarchy checks in this way.

If, for the moment, we change notation a little bit and let *CL* denote the class of the client, and *SE* the class of the server, then a more relaxed interpretation of the MR's precondition part can be written as $pre_{CL} \rightarrow pre_{SE}$; given that the former holds, then $pre_{CL} \wedge pre_{SE}$ can be derived using (T12). And this is the same condition that we have required for *cl-**PRE*** (see section 4.2). For both approaches, the only thing that matters is, if the client does expect and the server can provide a safe execution. The difference is that the MR uses $pre_{CL}$ as a guard for $pre_{SE}$: $pre_{SE}$ needs to hold only if the client expects the execution of the server method.

### 7.1.2   Postcondition checking

Again, assume that we deal with a *C*-client whose call is dynamically dispatched to a method $m_{SSC}$, and that the client precondition $pre_C$ holds. Then, applying a strong interpretation we get $pre_C \wedge post_{SC} \rightarrow post_C$, whence $pre_C \rightarrow (post_{SC} \rightarrow post_C)$ by (T14), and finally $post_{SC} \rightarrow post_C$ by modus ponens. Analogous steps lead to $post_{SSC} \rightarrow post_{SC}$. If $m_{SSC}$ works correctly, then $post_{SSC}$ holds, and we can derive $post_{SSC} \wedge post_{SC} \wedge post_C$. There is no problem with this, since we are working from inside out, i.e. from an extended to a reduced state space, and subclass postconditions can be stronger in a natural way.

Under the relaxed view on the MR we have $pre_{CL} \rightarrow (post_{SE} \rightarrow post_{CL})$. Given that both $pre_{CL}$ has been valid before execution of $m_{SE}$, we need $post_{SE}$ to hold afterwards, in order to be able to derive $post_{CL}$. $effConPost_{SE}$, on the other hand, (see (8), section 5.2) becomes equivalent to $view_{CL} \rightarrow post_{CL} \equiv client_{CL} \wedge pre_{CL} \rightarrow post_{CL}$, whence we can derive $post_{CL}$: *cl-**POST*** does not require $post_{SE}$ to be valid, unless *SE* is also the client class. This eventually can prove to be a kind of hazard, but there is a chance that the result that suffices for the client also does for further calculations of the program.

## 7.2   Inconsistent preconditions

**Example 1.** (for (P1), section 4.1). Assume that we have a method $m_C$ in class *C* with precondition $pre_C(x) := x > 0$, which is overridden in class *SC* by $m_{SC}$ with $pre_{SC}(x) := x < 0$ (possibly due to a typo or a misunderstanding). Further suppose that we have a call $cl_C.o_{SC}.m(-2)$.

Looking for the appropriate entry in *Table 3*, (i.e. for the one $pre_{SC}(-2) \equiv$ **true** and $pre_C(-2) \equiv$ **false**), column 3 tells us to reject the call for a client $cl_C$, contrary to the original percolation mechanism which uses $effPre_{SC} := pre_C \vee pre_{SC} \equiv$ **true** for decision and would lead to a surprising, but safe execution. Hence, our client aware reaction would solve this problem: as expected by the client, a call $cl_{SC}.o_{SC}.m(-2)$ would lead to an execution of $m_{SC}(-2)$.

**Example 2.** (for (P3), section 4.1). Consider the following call: $cl_C.o_{SC}.m(2)$, i.e. $m_{SC}(2)$, will result in an unexpected failure of $pre_{SC}(2)$, whereas the client expects a valid result. In this case we do not have an error on the client side (who uses a positive argument in accordance with its view, i.e. the precondition of $m_C$), but a hierarchy violation on method level specification. Therefore, at least in case of an error we have to check the hierarchy on the method level specification if we want to locate the error correctly. This kind of error goes unnoticed in the percolation and specification inheritance mechanisms: they both would perform an unsafe execution, whereas (due to column 2 of *Table 3*) the client conforming check rejects the above call, thereby giving a hint to an inconsistency in the precondition specifications involved.

### 7.3 Some open Issues

#### 7.3.1 The problem with unsafe executions

Though this may, of course, be contrary to what a client expects, we have excluded unsafe executions (i.e. the invocation of a method if its class specific precondition does not hold) in section 4.2 for good reasons. Concerning such unsafe executions, two alternative strategies can be considered for a call $cl_C.\,o_{SC}$: If $pre_C \equiv$ **true** and $pre_{SC} \equiv$ **false** then (a) execute $m_{SC}$ and let's see what happens; (b) execute $m_C$ instead of $m_{SC}$, since its precondition holds, i.e. redirect the call to the client class method. These two approaches may be regarded as variants of strong client conformance: the first a risky, and the other a safe one, that should fully correspond to the client's expectation. However, using approach (b) we also cannot know which effects will be caused in subsequent calculations by the proposed redirection.

#### 7.3.2 The problem with extended state space in subclasses

In section 7.1.1 we have shown that valid hierarchy checks as induced by a strict interpretation of the MR force subclass methods to fulfill a stronger precondition. Moreover, the MR requirement that a supertype precondition must imply that of an overriding method in a subclass is counterintuitive in the following sense: We must allow subclasses to specify constraints on their eventually extended state space. Such a precondition may naturally be stronger than that in a superclass, or – as an extreme scenario – may not be related to it at all. ([Find01] discusses a simple example (a>0 versus a>10) given in section 4.1 of [Karao99]. To our knowledge, the latter is the first hint on this kind of hierarchy errors and the problems they yield for traditional effective precondition construction.)

Current RAC machines do not perform hierarchy checks on developer written contracts, but instead use one of the percolation pattern variants to calculate the constraints used for runtime checks. As a consequence, the MR requirements are lifted from the developer written contracts on method level to the effective constraints. This may have the consequence that design errors become masked, get allocated incorrectly, or lead to a semantics not intended by developers in situations like the one in Example 3 below: Since $pre_T \vee (pre_T \wedge addedPre_S) \equiv pre_T$ by simple lattice algebra, the evaluation of $addedPre_S$ becomes irrelevant.

On the other hand, using percolation allows for strengthening preconditions in subclasses to get adequate assertions for the new fields in the extended state space: Such a strengthening cannot harm superclass clients, because of the structural properties of percolation pattern: If the client class precondition, say $pre_T$, for a method $m_T$ holds, then so does every disjunction containing $pre_T$. Therefore, subclasses of $T$ may change their precondition for $m$, and hence also strengthen it by adding additional requirements for new fields.

**Example 3.** With this simple example, which uses a natural precondition strengthening on the extended state space of a subclass, we demonstrate that also client conforming subtyping cannot handle all contracts in a desirable way. Assume $a$ stands for the amount a customer wants to withdraw from his or her account, and $c$ is a counter holding the limited number of withdrawals allowed for a single day. Then we might have e.g. $pre_C(a) := 0 \leq a \leq 1000$, which is overridden in class $SC$ with $pre_{SC}(a,c) := pre_C(a) \wedge c \leq maxDailyWithdrawals$.

Let $a = 200$, $c > maxDailyWithdrawals$ and consider the call $cl_C.o_{SC}.m(200)$. *s*-**PRE** would accept this call for both $cl_C$ and $cl_{SC}$ clients, and start an unsafe execution of $m_{SC}$. But what we probably really want is: The call $cl_{SC}.o_{SC}.m(200)$ should be rejected, because the second part of the precondition $pre_{SC}(a,c)$ is violated, and the call $cl_C.o_{SC}.m(200)$ should be accepted, because $cl_C$ does not know about a daily limit, and its precondition part is valid. Client conforming contract checking, aiming for safe executions, would reject both calls, quite the opposite of the simple percolation approach.

So it seems that configurations in real life software can be so diverse that there is no best specification paradigm; but, hopefully, there are some good ones. A more detailed analysis of the logical structure of the precondition, and to which class the variables involved do belong to, would maybe allow a more suitable decision. In the example above, the precondition failure is entirely located in the class frame of $SC$, and one might consider to let the call $cl_C.o_{SC}.m(200)$ execute $m_{SC}$ in this case. Perhaps the concepts of class frames and frame typestates [DeLine04] deserve consideration to learn if they can help here.

### 7.4 Related work

The topics discussed here have first been described in [Toth10]. Astonishingly, it seems that only one paper has been presented during the last three years that explicitly deals with the same kind of situations as I do in this paper: In [ReLea12], section 4.1 on Inconsistent Reasoning Problem, the authors give examples of what they call "masked precondition violation" and "unexpected contract violation", respectively. The first one leads to what I have called a surprising execution (see section 4.1 above), and the second one to a postcondition violation due to a surprising failure caused by strengthened constraints in a subclass (see section 5.1). The solution proposed in [ReLea12] is described as follows (on page 10): "To avoid the problems described above, one needs to avoid the overly-dynamic nature of contemporary RAC compilers, which results from their checking method specifications on the supplier side. By contrast, our approach, which we call client-aware interface specification checking, or CAISC, uses client-side checking to avoid these problems." Consequently, runtime checks are inserted at the site of method calls, whence the statically visible constraints can be checked. As already stated above, this technique solves the problem with surprising, but does not avoid unsafe executions. Moreover, it probably causes an insertion overhead.

# 8  Summary and conclusions

In this paper I have deliberately followed an important software engenieering principle: First define, *what* you want to have; then check if it is *achievable* at all, and if yes, decide *how* this will be done. Clearly, only the first of these three tasks has been tackled here. If the informal concepts put forward turn out to be reasonable ones, then we should find ways for their efficient implementation in order to benefit from their improved diagnosis possibilities, and should also build formal semantics.

I have proposed an alternative view on the basic mechanisms and concepts of runtime assertion checking. The considerations have been restricted to specifications of methods with pre- and postconditions, and class invariants as the classical backbone of software contracts. The search for the new approach has been triggered by various weaknesses that could be tracked down in two popular systems: Eiffel's Design by Contract, and specification inheritance used by JML and Spec#.

The difficulties with these approaches, and many tools using the same techniques, result from the insufficient handling of type-related information. Clearly, this kind of information is not an element of what is classically regarded as program state, and reflective abilities of programming languages and their usage have not been part of traditional mainstream investigation for runtime assertion checking. But as has been demonstrated, properties of the program state alone (more precisely: of the values in the attribute fields of instances) are not accurate enough to formulate contracts for the various client-server configurations in object-oriented programs in a proper way.

What is new in client conforming contracts is respecting the client's view for reasoning about the program state (determined by its static type) as a guard or filter on the way how runtime checks are done. This kind of client awareness is contrary to the server focused approach of percolation mechanisms commonly used: In the traditional approaches the client type is unknown, and the corresponding RACs handle all kinds of possible clients in the same way, based only on the dynamic type of the server. Therefore, the checks must be prepared (preconditions) and provide results (postconditions, invariants) for all possible cases, and in doing so become more unspecific than necessary, imposing undesired side effects. As we have shown, it needs only small syntactical modifications of the commonly used percolation pattern (see *Table 7*) to achieve important semantical changes in specifications, with considerable positive consequences on contract formulation and their runtime checking. Although client conforming subtyping offers a lot of advantages compared to other approaches, it is not the ultimate ratio (see Example 3 in section 7.3.2) and offers further opportunities for improvements.

|  | $\bigvee_{C \in supers(S)} (\bullet)$ |  | $\bigwedge_{C \in supers(S)} (\bullet)$ |  |
|---|---|---|---|---|
| **s-PRE** | $pre_C$ | s-**POST** | $post_C$ | Eiffel, most RACs |
|  |  | g-**POST** | $pre_C \rightarrow post_C$ | join composition, JML, specification inheritance |
| **cl-PRE** | $view_C \wedge pre_S \equiv$ $client_C \wedge pre_C \wedge pre_S$ | **cl-POST** | $view_C \rightarrow post_C \equiv$ $client_C \rightarrow (pre_C \rightarrow post_C)^{*)}$ | client conformant percolation |
| **ConRACPre** | $client_C \wedge pre_C \wedge inv_C$ $\wedge pre_S \wedge inv_S$ | **ConRACPost** | $client_C \rightarrow (pre_C \wedge \mathbf{old\_inv}_C \rightarrow post_C \wedge inv_C)^{*)}$ |  |

*Table 7*: Overview of percolation pattern variants

*) $client_C \wedge pre_C \rightarrow post_C \equiv client_C \rightarrow (pre_C \rightarrow post_C)$ by (T14)

In a nutshell, client conforming subtyping is maybe not the best solution, but it provides improvements over traditional RAC, simultaneously keeping things simple for developers:
- surprising and unsafe executions are ruled out;
- surprising failures of postconditions are ruled out;
- due to the 'client-filter' we do not have to bother with the extended state problem, since developers can define constraints in subclasses independent from their superclass constraints, i.e. strengthening or changing is possible to cope with the extended state space;
- the view of a specification as a contract between a client and a server is emphasized by neglecting classes other than the static type of the client and the dynamic type of the server, again due to the 'client-filter'; (Note that this should allow for an efficient implementation.)
- the asymmetry of the server focused approach of contemporary RAC mechanisms (due to neglecting the client's static type) is avoided by more carefully formulated cooperation contracts that both protect a client's view and also provide the necessary working conditions for the server method (see section 6.1);
- as other approaches and tools do, a variant of the percolation pattern is used, whence we automatically get the plug-in property for efficient constraints, but with a considerably more focused semantics.

The problem to be solved is how to make the static type of the client available to the server (where we, as usually, assume that the burden for performing the checks is imposed on the server method).

There exists a general problem in the area of software specification: Scanning the related literature we find a huge amount of papers on formal semantics, sophisticated logical calculi and verifiers. This is very interesting and a wonderful progress. But we must go further: There are only very few guidelines concerning aspects and problems that average software engineers are confronted with in their daily work; I believe that it is also very important to extend this short list (comprising e.g. [Mitch02], [Nunes04], and [Ruby06]) and to discuss the many facets of specification practice more thoroughly. This will help close the gap between the theory of software specification and software development in practice. I hope that this paper can prove to be a contribution to this goal.

## *Appendix: Some logical aspects of percolation pattern variants*

I cannot help but give a kind of rehabilitation: Not only are very large parts of the progress in the discussion of program specification and verification inconceivable without them, but Eiffel's DbC mechanism and specificationn inheritance also are – despite the weaknesses we have identified – highly reasonable approaches to behavioral subtyping: Both satisfy an adequate optimality criterion within their semantic requirements on what specifications have to accomplish, as shown below.

**1. Eiffel's percolation mechanism**. In section 3.3 the following relationships have been justified:

$$(pre_C \vee pre_{SC} \vee pre_{SSC}) \rightarrow effPre_{SSC} \qquad (A1)$$

$$effPost_{SSC} \rightarrow (post_C \wedge post_{SC} \wedge post_{SSC}) \qquad (A2)$$

Since we do not want to require more than is necessary, we may ask if $pre_C \vee pre_{SC} \vee pre_{SSC}$ is a least upper bound for $effPre_{SSC}$ (remember: $B \leq A := A \rightarrow B$). The answer is yes: Given that $effPre_{SSC}$ holds, we know that - due to the standard typing rules for object-oriented languages - $m_{SSC}$ is open for calls from superclass clients only. Thus, at least one of $pre_C$, $pre_{SC}$ or $pre_{SSC}$ must be fulfilled, i.e. $effPre_{SSC} \rightarrow (pre_C \vee pre_{SC} \vee pre_{SSC})$. Together with (A1) above we then get $effPre_{SSC} \equiv (pre_C \vee pre_{SC} \vee pre_{SSC})$. So *s-**PRE*** looks fine.

A similar argument shows the analogous property for *s-**POST***: $post_C \wedge post_{SC} \wedge post_{SSC}$ is a greatest lower bound for $effPost_{SSC}$, i.e. it provides as much as possible. Suppose there exists a lower bound $effPost'_{SSC}$ weaker than $post_C \wedge post_{SC} \wedge post_{SSC}$, i.e. $(post_C \wedge post_{SC} \wedge post_{SSC}) \rightarrow effPost'_{SSC}$. Since, by (A2) above, also the reverse implication must hold for $effPost'_{SSC}$, we get $effPost'_{SSC} \equiv (post_C \wedge post_{SC} \wedge post_{SSC})$.

Thus, Eiffel's percolation mechanism does the optimal work within its conceptual context (i.e. the semantic framework of requirements it imposes on contracts): The disjunction of preconditions is a least upper bound, and the conjunction of postconditions a greatest lower bound for what an overriding method in a subclass must accept and provide, respectively.

**2. Specification inheritance.** Though join composition does partly help with problem (P5) (see section 5.1.2), it is in a certain sense the best what we can get. This is the main reason why we present the following results, and also because there exist two prominent implementations: JML and Spec# (see [Leave06] and [LeiMü09]). The detailed theoretical background can be found in [LeNa06a] and [LeNa06b], two rather technical papers; a more accessible presentation is given in [Leave06], to which we adhere but translate to the terminology used here and provide only that level of generality which is needed for our considerations. (For a better orientation for the reader I use the original numbering of definitions and theorems of [Leave06].)

The technique of supertype abstraction validates reasoning about a dynamically dispatched method call, say $cl_{SC}.o_{SSC}$, using the specification of the static type of the navigation expression $cl_{SC}$. The idea behind is that objects of all subtypes $S$ of a type $T$ (including $T$ itself) can be treated uniformly. Supertype abstraction thus permits reasoning about programs that are open to the addition of new subtypes and, in this sense, is a modular reasoning technique.

**Definition 1.** (Join of method specifications) Let $\langle pre_S, post_S \rangle$ and $\langle pre_T, post_T \rangle$ be specifications of an instance method $m$, where $S$ is a subtype of $T$. Then their *join*, written as $\langle pre_S, post_S \rangle \sqcup \langle pre_T, post_T \rangle$, is the specification

$$\langle pre_T \vee pre_S, (\mathbf{old\_}pre_T \rightarrow post_T) \wedge (\mathbf{old\_}pre_S \rightarrow post_S) \rangle.$$

(Remember: We write **old_**pre to make explicit that precondition evaluation is to be understood on the attribute values before execution of the method specified.)

Remember that this is just join compositon as introduced at the end of section 5.1.2, now applied to class hierarchies: the "set of possible clients, each with its own specific expectations" for some subclass method $m_S$ now becomes the set of possible callers statically typed with an arbitrary supertype $T$ of $S$. Joining method specifications from subtypes with those inherited from supertypes is the vehicle for achieving specification inheritance in the calculation of the effective constraints as follows:

**Definition 2.** (Specification inheritance) Suppose $T$ has supertypes $supers(T)$. Then the effective specification of $T$ is a specification such that:
(a) For all methods $m$ in any of the types in $supers(T)$, the effective specification of $m$ in $T$ is the join of the specifications for $m$ in $T$ and all its proper supertypes:

$$effSpec_T := \bigsqcup \{ spec_U \mid U \in supers(T) \}$$

(b) The effective invariant of $T$ is the conjunction of all the invariants in $T$ and its proper supertypes:

$$effInv_T := \bigwedge \{ inv_U \mid U \in supers(T) \}$$

**Theorem 1.** Let $S$ be a subtype of $T$, and $\langle pre_S, post_S \rangle$ and $\langle pre_T, post_T \rangle$ be specifications of an instance method $m$ in $S$ and $T$, respectively. Then $\langle pre_S, post_S \rangle \sqsupseteq \langle pre_T, post_T \rangle$ if and only if the following two conditions hold:

(1) $pre_T \rightarrow pre_S$, and (2) $\mathbf{old\_}pre_T \rightarrow (post_S \rightarrow post_T)$

**Theorem 2.** Let $S$ be a subtype of $T$, and $\langle pre_S, post_S \rangle$ and $\langle pre_T, post_T \rangle$ be specifications of an instance method $m$. Then

$$\langle pre_S, post_S \rangle \sqcup \langle pre_T, post_T \rangle \sqsupseteq \langle pre_T, post_T \rangle.$$

(Note that conditions (1) and (2) of Theorem 1 define the relaxed plug-in match (see end of section 3.2), which has been proven to be a most general reuse ensuring specification match in Theorem 7 of [CheChe00].)

The most important notion of behavioral subtyping, which corresponds to Liskov and Wing's constraint-based definition [LisWin94, p.1823], is given in

**Definition 5.** (Strong behavioral subtype) Let $S$ be a subtype of $T$. Then $S$ is a strong behavioral subtype of $T$ if and only if:
(a) For all instance methods $m$ in $T$, the method specification for $m$ in $S$ refines that of $m$ in $T$;
(b) The instance invariant of $S$ implies the instance invariant of $T$ for objects of type $S$.

**Theorem 3** Let $S$ be a subtype of $T$. Then the effective specification of $S$ is a strong behavioral subtype of the effective specification of $T$.

A result analogous to that for Eiffel's percolation mechanism holds: We show that the join of method specifications is their least upper bound in the refinement ordering. Theorem 2 states the upper bound property: Trivially, $pre_T \rightarrow pre_T \vee pre_S$. Assume that $pre_T$ holds; from $(pre_T \rightarrow post_T) \wedge (pre_S \rightarrow post_S)$, we get $pre_T \rightarrow post_T$, and applying modus ponens, $post_T$ can be derived. Now by Theorem 1 $\langle pre_S, post_S \rangle \sqcup \langle pre_T, post_T \rangle \sqsupseteq \langle pre_T, post_T \rangle$.

To show the least upper bound property, assume that

$$\langle pre, post \rangle \sqsupseteq \langle pre_0, post_0 \rangle \wedge \langle pre, post \rangle \sqsupseteq \langle pre_1, post_1 \rangle. \tag{A3}$$

Thus, $(pre_0 \rightarrow pre) \wedge (pre_1 \rightarrow pre) \equiv (pre_0 \vee pre_1 \rightarrow pre)$, by (T17). Furthermore, from (A) we also get $(post \rightarrow post_0) \wedge (post \rightarrow post_1) \equiv (post \rightarrow post_0 \wedge post_1)$ by (T16). We have

$[post \rightarrow (post_0 \wedge post_1)] \rightarrow [(pre_0 \vee pre_1) \wedge post \rightarrow (post_0 \wedge post_1)]$, by (T19), and

$(post_0 \wedge post_1) \rightarrow (pre_0 \rightarrow post_0) \wedge (pre_1 \rightarrow post_1)$, by (T18).

Transitivity of implication gives

$(pre_0 \vee pre_1) \wedge post \rightarrow (pre_0 \rightarrow post_0) \wedge (pre_1 \rightarrow post_1)$,

and by (T14) and Theorem 1 we finally get $\langle pre, post \rangle \sqsupseteq \langle pre_0, post_0 \rangle \sqcup \langle pre_1, post_1 \rangle$.

x